\documentclass[twocolumn,prd,nofootinbib,aps,prl,floats,floatfix,amsmath,amssymb,longbibliography,secnumarab
ic,superscriptaddress,preprintnumbers]{revtex4} %
\usepackage[final]{graphicx}
\usepackage{hyperref}
\usepackage{amsmath}
\usepackage{bbm}
\usepackage{bm}
\usepackage{amsfonts}
\usepackage{amssymb}
\usepackage{latexsym}
\usepackage{mathtools}
\usepackage{graphicx}
\usepackage[english]{babel}
\usepackage{multirow}
\usepackage{float}
\usepackage{url}
\usepackage{slashed}
\usepackage{xcolor} 
\usepackage[utf8]{inputenc}
\usepackage{stmaryrd} 
\usepackage{enumitem}
\usepackage{hyperref}
\usepackage{cleveref}
\usepackage{siunitx}
\usepackage{verbatim}

\newcommand{\be}{\begin{equation}}
\newcommand{\ee}{\end{equation}}
\newcommand{\ba}{\begin{array}}
\newcommand{\ea}{\end{array}}
\newcommand{\bea}{\begin{eqnarray}}
\newcommand{\eea}{\end{eqnarray}}

\newcommand{\besub}{\begin{subequations}}
\newcommand{\eesub}{\end{subequations}}

\renewcommand{\eqref}[1]{Eq.~(\ref{eq:#1})}

\def\beq{\begin{equation}}
\def\eeq#1{\label{#1}\end{equation}}
\def\eeqn{\end{equation}}
\def\beqa{\begin{eqnarray}}
\def\eeqa#1{\label{#1}\end{eqnarray}}
\def\eeqan{\end{eqnarray}}
\def\CR{\nonumber \\ }
\def\leqn#1{(\ref{#1})}

\newcommand{\centeron}[2]{{\setbox0=\hbox{#1}\setbox1=\hbox{#2}\ifdim
		\wd1>\wd0\kern.5\wd1\kern-.5\wd0\fi \copy0
		\kern-.5\wd0\kern-.5\wd1\copy1\ifdim\wd0>\wd1
		\kern.5\wd0\kern-.5\wd1\fi}}
\newcommand{\ltap}{\>\centeron{\raise.35ex\hbox{$<$}}
	{\lower.65ex\hbox{$\sim$}}\>}
\newcommand{\gtap}{\>\centeron{\raise.35ex\hbox{$>$}}
	{\lower.65ex\hbox{$\sim$}}\>}

\newcommand{\lsim}{\mathrel{\ltap}}

\begin{document}

\title{Z-portal Continuum Dark Matter }

\author{Csaba Cs\'aki}
\affiliation{Department of Physics, LEPP, Cornell University, Ithaca, NY 14853, USA}

\author{Sungwoo Hong}
\affiliation{Department of Physics, LEPP, Cornell University, Ithaca, NY 14853, USA}
\affiliation{Department of Physics, The University of Chicago, Chicago, IL 60637 , USA }
\affiliation{Argonne National Laboratory, Lemont, IL 60439, USA}

\author{Gowri Kurup}
\affiliation{Department of Physics, LEPP, Cornell University, Ithaca, NY 14853, USA}
\affiliation{Department of Physics, University of Oxford, Parks Rd, Oxford OX1 3PJ, United Kingdom}

\author{Seung J. Lee}
\affiliation{Department of Physics, Korea University, Seoul, 136-713, Korea}

\author{Maxim Perelstein}
\affiliation{Department of Physics, LEPP, Cornell University, Ithaca, NY 14853, USA}

\author{Wei Xue}
\affiliation{Department of Physics, University of Florida, Gainesville, FL 32611, USA}

\begin{abstract}

We examine the possibility that dark matter (DM) consists of a gapped continuum, rather than ordinary particles. A Weakly-Interacting Continuum (WIC) model, coupled to the Standard Model via a Z-portal, provides an explicit realization of this idea. The thermal DM relic density in this model is naturally consistent with observations, providing a continuum counterpart of the ``WIMP miracle". Direct detection cross sections are strongly suppressed compared to ordinary Z-portal WIMP, thanks to a unique effect of the continuum kinematics. Continuum DM states decay throughout the history of the universe, and observations of cosmic microwave background place constraints on potential late decays. Production of WICs at colliders can provide a striking cascade-decay signature. We show that a simple Z-portal WIC model provides a fully viable DM candidate consistent with all current experimental constraints.

\end{abstract}

\maketitle

\section{Introduction}
\label{sec:intro}

Strong evidence from cosmology and astrophysics points to the existence of dark matter (DM), which cannot be made up of standard model~(SM) particles. Many viable DM models 
have been proposed. Among these, the weakly-interacting massive particle~(WIMP) DM~\cite{Lee:1977ua} has been the leading paradigm for decades. WIMPs arise from well-motivated theoretical extensions of the SM, such as the neutralino DM in supersymmetry \cite{Jungman:1995df}. The WIMP paradigm naturally reproduces the observed DM density; moreover, the relic abundance of WIMPs is insensitive to the initial conditions of the Universe due to the thermal equilibrium between the DM and SM gases in the early Universe. However, experimental searches for non-gravitational signatures of WIMPs have not found any positive evidence yet, leading to increasingly tight constraints on this idea~\cite{Bertone:2004pz}.
For example, scalar Z-portal DM, one of the simplest WIMP models, is ruled out by several orders of magnitude~\cite{Escudero:2016gzx} by direct detection experiments 
\cite{Aprile:2018dbl,Akerib:2016vxi,Cui:2017nnn}. 

\begin{figure}
	\begin{center}
	 \includegraphics[width=9.5cm]{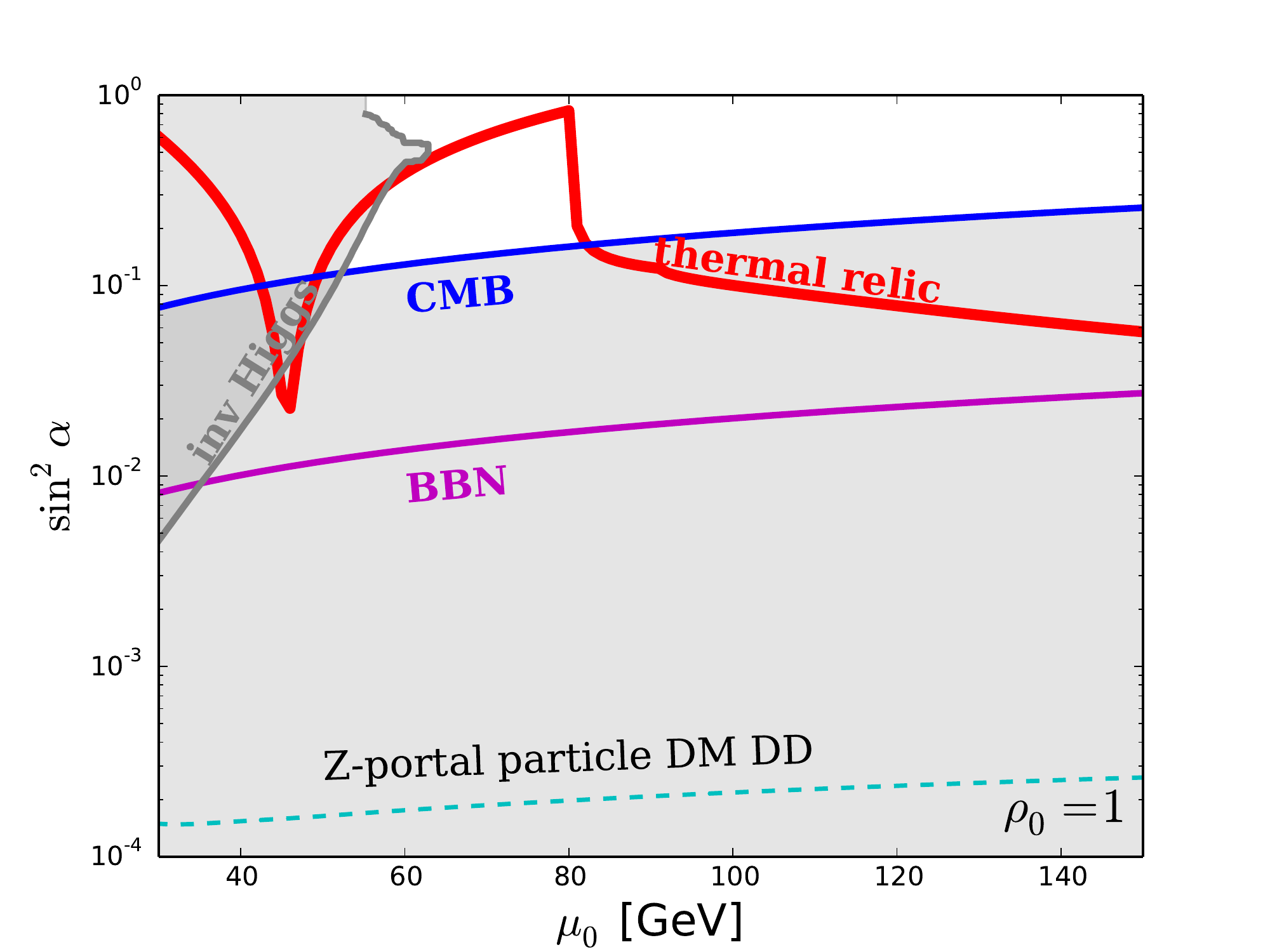} 
	\end{center}
	\caption{Parameter space of the Z-portal WIC, with $\rho_0=1$. Red curve: thermal relic consistent with observations
		\cite{Ade:2015xua}. Space below the blue curve is ruled out by the CMB constraint~\cite{Slatyer:2016qyl,Slatyer:2015jla}, while space below the magenta curve is also ruled out by the BBN constraint on electromagnetic energy injection~
		\cite{Reno:1987qw,Kawasaki:1994sc,Cyburt:2002uv,Kawasaki:2004qu,Hisano:2009rc,Henning:2012rm}. Space above the gray curve is ruled out by the LHC constraint on exotic Higgs decays (with $m_\chi=500$~GeV). For comparison, the bound of DM direct detection for Z-portal {\it particle} DM is shown as the dotted cyan curve.  
	}
	\label{fig:Zportal} 
\end{figure}

In this letter, together with a companion paper~\cite{Csaki:2021gfm}, we propose a novel framework for DM, based on quantum field theories with a gapped continuum spectrum. Rather than ordinary particles, in such models DM consists of a mixture of states with a continuous mass distribution above a certain ``gap" scale $\mu_0$. Within this framework, we focus on the ``weakly-interacting continuum" (WIC) scenario, where the gap  $\mu_0\sim 100$~GeV is near the electroweak scale and the continuum DM interacts with the SM via weak interactions. We argue that such a WIC model will maintain all the attractive features of WIMPs, including the natural consistency with the observed DM abundance and insensitivity of relic density to initial conditions. At the same time the continuum nature of the DM leads to striking new phenomenological features.\footnote{Our model shares some features with dynamical DM models proposed in~\cite{Dienes:2011ja, Dienes:2011sa,Dienes:2012yz,Dienes:2012cf,Dienes:2013xya,Dienes:2014via,Dienes:2014bka,Boddy:2016fds,Curtin:2018ees,Dienes:2019krh}, but the two frameworks also have important differences in both model-building and phenomenological predictions. For a detailed comparison, see Ref.~\cite{Csaki:2021gfm}. Also, in~\cite{vonHarling:2012sz,Katz:2015zba,Chaffey:2021tmj}, a continuum is used as the mediator to the DM, while here the continuum is the DM itself.}  In particular, the distinct kinematics of low-energy scattering of continuum states leads to strong suppression of direct detection rates, reviving the possibility of Z-portal DM.
In this paper we focus on the Z-portal model as a simple and explicit realization of WIC paradigm. The second unique phenomenological feature is that  throughout the history of the Universe, DM states continuously decay via DM$(\mu_1) \to $DM$(\mu_2) + {\rm SM}$ processes. Late decays of this kind may leave observable effects in the cosmic microwave background (CMB) due to energy deposition during and after the recombination epoch. 
At collider, production of heavy DM states may lead to spectacular signatures as they cascade-decay to invisible states near the gap and SM particles with characteristic multiplicities and spectra.

Gapped continuum theories have been used in various contexts in particle and condensed matter physics, including applications of unparticles~\cite{Georgi:2007ek} to Higgs physics and the hierarchy problem~\cite{Stancato:2008mp,Falkowski:2008yr,Bellazzini:2015cgj,Csaki:2018kxb}, string theory~\cite{Gubser:2000nd}, the fractional quantum Hall effect~\cite{sachdev2007quantum,Fradkin:1991nr} or the 2D Ising model~\cite{McCoy:1978ta}.
The theoretical framework for quantitative studies of continuum DM is presented in detail in the companion paper~\cite{Csaki:2021gfm}. There we described how to construct the Hilbert space of gapped continuum theories, gave formulae for calculating rates of scattering and decay involving continuum DM states, and discussed how to treat their equilibrium and non-equilibrium thermodynamics. We also presented an explicit model of Z-portal WIC, based on a warped 5D soft-wall geometry~\cite{Randall:1999ee, Cabrer:2009we}, 
and applied our formalism to calculate the DM relic density in this model. After a brief review of the Z-portal WIC model, this letter focuses on its phenomenology. The main results of our analysis are summarized in \cref{fig:Zportal}: it is demonstrated that the Z-portal WIC model can reproduce the observed DM density while being fully consistent with current experimental constraints, including direct detection, CMB, and collider data.

\section{Z-portal WIC model} 

The continuum DM is described by a ``generalized free field" $\Phi$, with an effective Lagrangian 
\begin{equation}
S = \int \frac{d^4 p}{(2\pi)^4} \; \Phi^\dagger (p) \Sigma (p^2) \Phi (p).
\label{eq:action_scalar_continuum}
\end{equation} 
Here $\Phi$ is a complex, gauge-singlet Lorentz scalar. The spectral density $\rho$ can be obtained as 
\bea
\rho (\mu^2) = -2 \; {\rm Im} \frac{1}{\Sigma (\mu^2)}.
\label{eq:rho_scalar}
\eea
In a gapped continuum theory, $\rho$ has continuous support above the gap scale $\mu^2_0$, and vanishes below that scale. Physically, the continuous parameter $\mu$ plays the role of the DM state mass. The singly-excited sector of the Hilbert space consists of states $|{\bf p}, \mu^2\rangle$, and the function $\rho(\mu^2)$ is the density of states with respect to $\mu^2$. If the continuum arises from a known 5D theory, the spectral density can be calculated. The behavior of $\rho$ in the vicinity of the gap scale is especially important for DM phenomenology, since during thermal freeze-out and throughout the subsequent evolution of the universe up to our own epoch, most of the DM states are clustered near the gap (see appendix~\ref{app:Decay}). It was shown in~\cite{Csaki:2021gfm} that in a broad range of models, the spectral density near the gap takes the generic form 
\begin{equation}
\rho(\mu^2) =  \frac{\rho_0}{\mu_0^2} \,\left(\frac{\mu^2}{\mu_0^2}-1\right)^{1/2},
\label{eq:rho0}
\end{equation}
where $\rho_0$ is an order-one dimensionless constant. As discussed in~\cite{Csaki:2021gfm}, the value of $\rho_0$ is model dependent and is fixed by the normalization of a brane-to-brane propagator in the corresponding 5D theory. In \cref{fig:Zportal} we show constraints for $\rho_0 =1$ as a benchmark. To illustrate the impact of varying this parameter, the constraint plot for $\rho_0=2\pi$ is included in the supplementary material.  
  
To couple $\Phi$ to the SM sector, a new complex scalar $\chi$ is introduced, which is an $SU(2)$ doublet and carries a $U(1)_Y$ charge, $Y =- \frac{1}{2}$. Both $\Phi$ and $\chi$ are odd under a $Z_2$ symmetry responsible for DM stability, while all SM fields are even. The field $\chi$ has a canonical kinetic term and $m_\chi^2>0$. An interaction term 
\begin{equation}
   {\cal L}_{\rm int} = - \lambda \, \Phi \, \chi \, H  \, + \,  {\rm h.c.}
   \label{eq:Lint0}
\end{equation}
leads to mass mixing of $\chi$ with the continuum after electroweak symmetry breaking. Assuming $m_\chi\gg \mu_0$ and
integrating out $\chi$ yields an effective Lagrangian coupling the continuum to SM:
\begin{align}
{\cal L}_{\Phi \text{-} Z,W} &= \sin^2 \alpha \left[ - \frac{i \sqrt{g^2+g'^2}}{2} 
\left( \partial_\mu \Phi^\dagger \Phi - \Phi^\dagger \partial_\mu \Phi  \right)  Z^\mu 
\right. 
\nonumber \\
&
\left.
+ \frac{g^2+g'^2}{4} \Phi^\dagger \Phi Z_\mu Z^\mu 
+ \frac{1}{2} g^2 \Phi^\dagger \Phi W^+_\mu W^{- \mu} \right].  
\label{eq:4D_Phi_ZW}
\end{align}
Here $ g$ and $g'$ are the SM $SU(2)_L$ and $U(1)_Y$ gauge couplings, and the mixing angle $\alpha$ is given by 
\begin{equation}
\tan 2 \alpha = \frac{ \sqrt{2} \, \lambda \, v }{  m_\chi^2 - \mu^2 }  \ .
\end{equation}
These interactions are responsible for WIC annihilation and decay. Note that the total number of the DM continuum states is conserved, even though they can decay into each other. The physics of WIC is described in terms of just three new parameters (the gap scale $\mu_0$, the mixing angle $\alpha$, and the spectral density normalization $\rho_0$), making for a simple and predictive model.

\section{Boltzmann equations and thermal relic} 

A dilute gas of continuum DM states is described by the occupation number $f({\bf p} , \mu)$. 
In thermal equilibrium, the occupation number follows the usual Fermi-Dirac or Bose-Einstein distribution, with particle mass $m$ replaced by the continuum state mass $\mu$. 
The number density of continuum states is given by 
\begin{equation}
   n = \int \frac{ {\rm d} \, \mu^2} {2 \pi } \nu(\mu^2)\,,
\end{equation}
where
\beq
\nu(\mu^2) = \rho ( \mu^2) \int \frac{ {\rm d}^3 p } { ( 2 \pi )^3  } f( {\bf p} , \mu) 
\eeq{nu_def}
is the DM mass distribution function.
When considering the freeze-out of continuum annihilation to the SM,
the Boltzmann equation for the continuum has the usual form (see detailed derivation in~\cite{Csaki:2021gfm}), 
\begin{equation}
  \frac{\partial n}{\partial t} + 3 H \, n = - \langle \sigma v \rangle (n^2 - n_{\rm eq}^2)\,, 
   \label{eq:boltz}
\end{equation}
where $n_{\rm eq}$ is the number density in thermal equilibrium. The thermal relic density for WIC is the same as for WIMP with the same thermally-averaged cross section $\langle \sigma v \rangle$, but for WIC $\langle \sigma v \rangle$ includes averaging over $\mu^2$. Taking into account the Boltzmann factor and the form of the spectral density near 
the gap, \cref{eq:rho0}, it is easy to see that the continuum distribution at temperatures below $\mu_0$ is localized near the gap scale.  The WIC relic density is then the same as for a WIMP of mass $m=\mu_0$ and interactions in \cref{eq:4D_Phi_ZW}, up to small corrections of order $\mathcal{O}( T^2 / \mu_0^2 )$. Note that the thermal relic density does not depend on $\rho_0$, since the normalization of the spectral density cancels out in the $\langle \sigma v \rangle$. 

The dominant WIC annihilation process during freeze-out depends on the gap scale $\mu_0$. In the regime 
of $\mu_0 < m_W$, the dominant channel is $ \Phi \, \Phi^* \to Z \to f  \bar{f}$. This process is p-wave, since
the parity of the initial state is $-1$. For a larger gap scale, $\mu_0 \gtrsim m_{W,Z}$, s-wave annihilations into pairs of vector bosons becomes relevant, $ \Phi \, \Phi^* \to W^+ W^-$ and $\Phi \, \Phi^* \to Z Z$. In \cref{fig:Zportal}, the red curve corresponds to 
the observed DM relic abundance. The required value of the mixing angle $\sin ^2 \alpha$ drops near the Z-pole, $\mu_0 \simeq m_Z/2$, where the cross section is enhanced by the Z-pole resonance. The decrease of the required $\sin^2\alpha$ above $m_W$ is due to opening of new annihilation channel. Note that we do not include the WIC annihilations to one on-shell and one off-shell W, which may give order-one corrections to the total annihilation rate in a narrow region of $\mu_0$ just below $m_W$.

\section{Late decay and CMB constraints}

One of the main distinguishing features of WIC DM compared to ordinary particle DM is that 
throughout the history of the universe, the continuum states are continuously decaying to
the SM particles, $\Phi(\mu_1) \to \Phi(\mu_2) + {\rm SM}$. These processes are always unavoidable   even in the presence of an exact $Z_2$ symmetry, due to the off-diagonal couplings among the continuum states such as in the interactions in \cref{eq:4D_Phi_ZW}. 
These decays keep dumping energy into the SM sector with $E \simeq \mu_1 - \mu_2$, and 
can leave imprints during and after 
the recombination epoch. 

\begin{figure}
	\begin{center}
		\hskip-0.5cm \includegraphics[width=8.8cm]{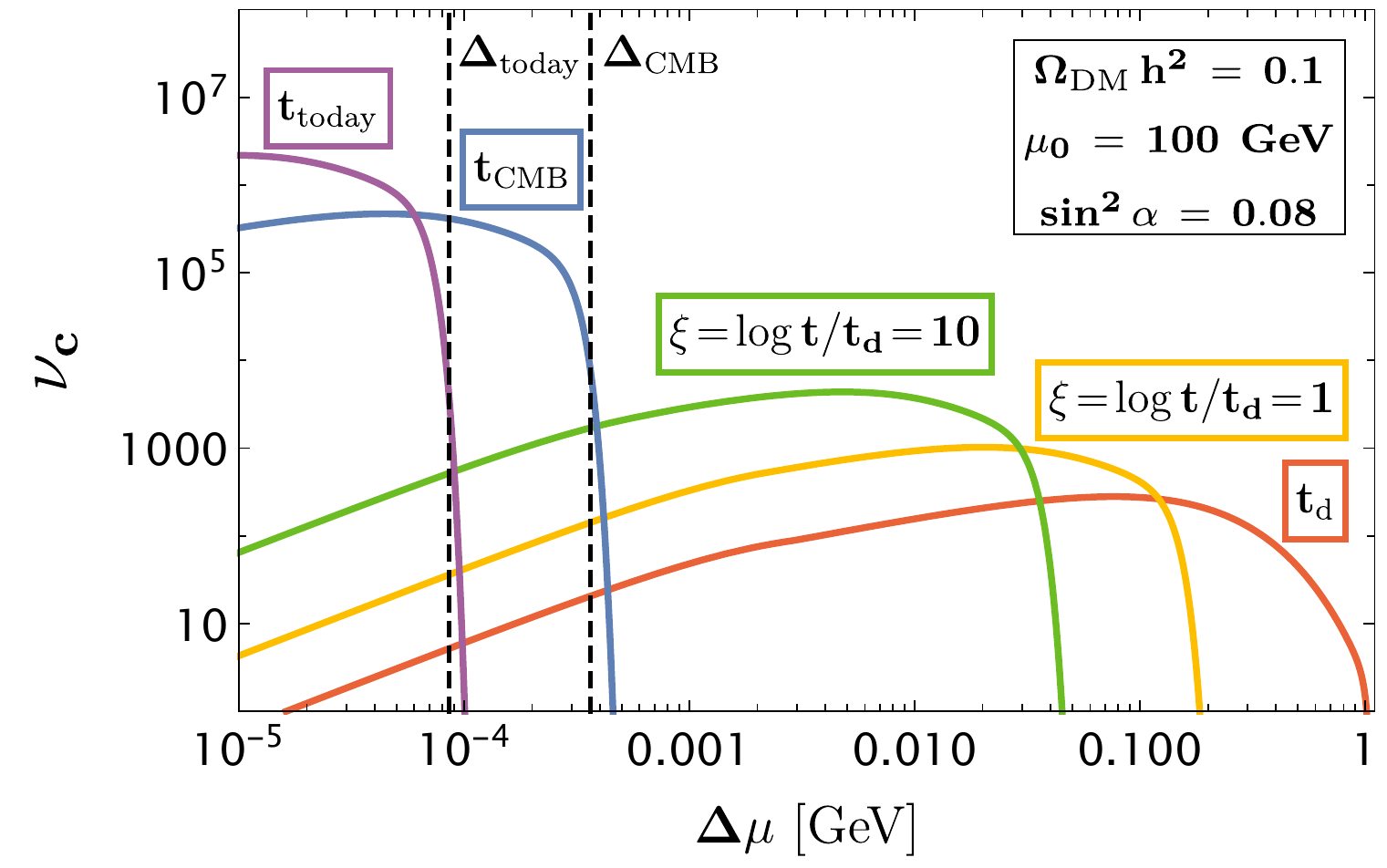} 
	\end{center}
	\caption{Time evolution of the DM comoving number density $\nu_c$, as a function of mass above the gap $\Delta\mu=\mu-\mu_0$, after its kinetic decoupling from the SM at time $t_d$. 
		The black lines indicate $\Delta_\mathrm{CMB}$ and $\Delta_\mathrm{today}$, found analytically by setting $\Gamma = H$ at photon decoupling and today, respectively. The y-axis units are arbitrary. See appendix~\ref{app:Decay} for details.
	}
	\label{fig:decay_log} 
\end{figure}

The evolution history of the continuum states can be summarized as follows. Before the freeze-out they are in thermal and chemical equilibrium. After the WIC freeze-out, which occurs at $T\sim \mu_0/10$, the quasi-elastic scattering $\Phi + {\rm SM} \leftrightarrow  \Phi + {\rm SM}$ and (inverse) decays $\Phi \leftrightarrow  \Phi + {\rm SM}$ maintain kinetic 
equilibrium of the continuum with the SM particles and chemical equilibrium among the continuum states themselves. This forces the mass distribution of the continuum states to peak closer to the gap scale $\mu_0$ as the SM plasma temperature decreases. At temperature $T$, a typical DM state has mass $\mu$ such that $\Delta\mu\equiv\mu-\mu_0\sim T$. Eventually, the rate of these processes falls below the Hubble rate, the WICs fully decouple from the SM, and the chemical equilibrium among the continuum states can no longer be maintained. For typical parameters in the Z-portal WIC model, this decoupling occurs at $T_d\sim 0.1-1$ GeV. After decoupling, the WIC mass distribution continues to evolve due to out-of-equilibrium WIC decays $\Phi \to  \Phi + {\rm SM}$, with typical masses tending ever closer to the gap scale. This evolution is illustrated in \cref{fig:decay_log}, which is based on a numerical solution of the Boltzmann equation 
(see supplementary material). An upper bound on the DM mass $\mu$ at time $t$ in this epoch can be estimated by equating the total decay rate of $\Phi(\mu)$ with the Hubble rate $H(t)$, since heavier DM states would have already decayed. This estimate agrees well with the numerical results, see \cref{fig:decay_log}. A typical DM mass at any given time is within a factor of a few from this bound, which implies that each DM state on average undergoes ${\cal O}(1)$ decays per Hubble time.       

In the Z-portal WIC model, the DM states decay through an off-shell Z, 
$\Phi(\mu_1) \to \Phi(\mu_2) + Z^* \to \Phi(\mu_2) +  f \bar{f}$. The rate of this decay (integrated over $\mu_2$) is given by 
\begin{align}
\Gamma ( \Phi \to \Phi + f \bar{f} ) 
& = \frac{  16  \sqrt{2}\, \rho_0  } { 15 \times 9009 \, \pi^4 }  \sin^4 \alpha \frac{ (g^2 + {g'}^2)^2}{ m_Z^4}
\nonumber \\
&  \hskip-2.5cm \times (g_A^2 + g_V^2 )  \,  (\Delta \mu )^5 \,  \left( \frac{\Delta \mu} {\mu_0} \right)^{3/2}   
\equiv \Gamma_0 \, \left( \frac{\Delta\mu}{\mu_0}\right)^{13/2}\, ,
\label{eq:decaytoee}
\end{align}
where $\Delta \mu = \mu_1 - \mu_0$, and $g_{A,V}$ are the SM Z couplings. The strong dependence on the DM mass arises from three-body phase space $\propto (\Delta \mu )^3$, the matrix element-squared $\propto (\Delta \mu )^2$, and final state spectral density integration 
$\propto  (\Delta \mu /\mu_0 )^{3/2}$. Here we assumed $\Delta\mu\gg m_f$; the rate rapidly drops to zero near the kinematic threshold $\Delta\mu=2m_f$. 

For typical WIC parameters, $\Delta\mu$ drops below muon and pion thresholds soon after kinetic decoupling, while the decays to $e^+e^-$ and the three neutrino flavors continue at late times. The first of these decays is potentially problematic for phenomenology, since it injects EM energy which can reionize hydrogen atoms after recombination. For example, if $\Delta\mu=2$ MeV soon after recombination, electrons of typical kinetic energy $\sim$ MeV would be produced, each of which can reionize $\sim 10^5$ H atoms. Given that $\rho_{\rm DM}/\rho_H\approx 5$ and that each DM state undergoes ${\cal O}(1)$ decays per Hubble time as explained above, the hydrogen would be quickly completely reionized, in gross conflict with CMB observations. Even decays close to kinematic threshold are ruled out. To avoid this constraint, we require that decays into $e^+e^-$ pairs become kinematically impossible at or before the CMB decoupling time $t_{\rm CMB}$. We conservatively estimate $\Delta\mu_{\rm CMB}$ using $\Gamma=H(t_{\rm CMB})$ (where $\Gamma$ is the total DM decay rate including neutrino final states), and require $\Delta\mu_{\rm CMB}< 2m_e \approx 1$ MeV. (In the supplementary material, we show that this simple estimate is supported by a more detailed analysis using numerically calculated DM mass distribution.) Another constraint arises from the photo-dissociation of nuclei due to electromagnetic energy injected by DM decays, which can change their primordial abundances and spoil the successful predictions of Big-Bang Nucleosynthesis (BBN)~\cite{Reno:1987qw,Kawasaki:1994sc,Cyburt:2002uv,Kawasaki:2004qu,Hisano:2009rc,Henning:2012rm}. The photo-dissociation 
becomes relevant when the temperature $T < 0.5 \, {\rm keV}$, which gives a weaker bound than the CMB.
The CMB and BBN constraints give {\it lower} bounds on the mixing angle $\sin\alpha$, shown by the blue and magenta curves in \cref{fig:Zportal}.      

After recombination, DM states continue to decay via the neutrino channel. Neutrinos from such decays could in principle be detected, but the predicted flux is well below the current constraints~\cite{Aprile:2020tmw,Bays:2011si,Agostini:2020lci,Agostini:2020mfq}.

\section{Direct Detection}

Direct detection signature of WIC is due to a scattering process DM$(\mu)+N\to$ DM$(\mu^\prime)+N$, where $N$ is a target nucleus. Since $\mu$ and $\mu^\prime$ are continuous variables, the kinematics of this process in the WIC model is quite distinct from either elastic or inelastic particle DM. This unique kinematics leads to a strong suppression of the direct detection rate, which is one of the most striking features of such models, and allows for a viable Z-portal WIC model consistent with current bounds.  

The direct detection rate for the continuum DM is 
\begin{equation}
   \frac{ {\rm d} R } {{\rm d} E_R} =  N_T \, \int \frac{d\mu^2}{2\pi} \,\nu_0(\mu^2)\,\int {\rm d}^3 v \, f( v ) 
          \frac{ {\rm d} \sigma  } {{\rm d} E_R} \, v 
\end{equation}
where $E_R$ is the recoil energy of the nucleus, $N_T$ the number of nuclei per unit mass, $\nu_0$ is the current WIC mass distribution function, and $f(v)$ is the local DM velocity distribution. We use the approximation
\begin{equation}
\nu_0 = \frac{\rho_\odot}  { \mu_0} \,\delta(\mu^2-\mu_1^2)\,,
\end{equation}
where $\rho_\odot = 0.3~{\rm GeV/cm^3}$ is the local DM density, and $\mu_1$ is found by equating the WIC decay rate with today's Hubble scale. (For typical parameters, $\mu_1-\mu_0 \sim 100-300$~keV.) We assume the Maxwell-Boltzmann distribution for the WIC DM with a cutoff at the escape velocity $v_{\rm esc} = 600~{\rm km / s}$. Since the DM state masses today are clustered near the gap scale with a small relative spread, the velocity distribution is  approximately mass-independent. 
The differential cross section is  
\begin{align}
   \frac{ {\rm d} \sigma } { {\rm d} E_R} 
   &= 
       \int_{\mu_0^2}^{\mu_{\rm max}^2}
      \frac{ {\rm d} \mu^{\prime 2} } { 2 \pi}  
      \, \rho( \mu^{\prime 2}) \,  \frac{ m_N \, \sigma_p } { 2 \mu_{\Phi n}^2 \, v^2 } 
      \nonumber
      \\
&
      \left[ \frac{  f_p Z + f_n ( A-Z) } {f_p} \right]^2
      F_N^2(q)  \ .
\label{eq:dsigmadER}
\end{align}
The DM-proton (or neutron) cross section is   
\begin{equation}
   \sigma_p = \frac{ (g^2 + g'^2)^2\, \sin^4 \alpha \, \mu_{\Phi n}^2  \, f_p^2 }{  16 \pi  m_Z^4 }   \ .
\end{equation}
Here $\mu_{\Phi n}$ is the reduced mass of the DM state and proton or neutron, and $F_N(q)$ is the nucleus form factor~\cite{Lewin:1995rx}. 
The coupling strength to nucleon  
$f_p$ ($f_n$) is given by the vector current matching from the quark to nucleon, 
$f_n = b_u + 2 b_d = -\frac{1}{4}$ and $f_p = 2 b_u + b_d = \frac14 - \sin^2 \theta_w$.
The incoming DM state with mass $\mu_1$ can be down-scattered or up-scattered. Using energy-momentum conservation, the maximum accessible mass for the outgoing state is 
\beq
\mu_{\rm max} = \mu_1 + q v - \frac 12 \frac{q^2} { \mu_{\Phi N} }\,,
\eeq{range}
where $q = \sqrt{2 m_N E_R} $ is the exchanged momentum. Since $\mu_1-\mu_0\ll \mu_0$ and $v\ll 1$, only a narrow range of DM states are kinematically accessible, suppressing the cross section by  
\begin{equation}
\int_{\mu_0^2}^{\mu_{\rm max}^2}  \frac{d\mu^{\prime 2}}{2\pi} \;  \rho (\mu^{\prime 2}) \sim \left( \frac{\Delta \mu}{\mu_0} \right)^{3/2}\,\sim {\cal O} ( 10^{-6} - 10^{-7}) .
\end{equation}

The direct detection bound was obtained by comparing the predicted number of events in the Z-portal WIC model  
with the corresponding particle DM prediction, and recasting the bounds from the XENON1T experiment~\cite{Aprile:2018dbl}. 
Thanks to the strong continuum kinematic suppression, XENON1T limits do not impose a constraint for $\rho_0=1$. (Direct detection constraints become relevant for larger $\rho_0$; see supplementary material.) For comparison, the limit on {\it particle} scalar Z-portal DM is shown by the dotted cyan curve in \cref{fig:Zportal}.

\section{Other constraints}
\label{sec:others}

Indirect detection does not currently constrain the WIC model. The kinematics and the cross section of DM annihilation in today's halos is virtually identical to usual particle DM of mass $\mu_0$. For Z-portal DM with $\mu_0<m_W$, the dominant channel is $ \Phi \Phi^* \to Z^* \to f \bar{f}$, and the rate is p-wave suppressed. For $\mu_0>m_W$, DM can annihilate via s-wave into $WW$. The rate is below the current bounds from FermiLAT~\cite{Fermi-LAT:2016uux}.

At colliders, continuous DM states can be pair-produced 
through virtual $Z$ exchange. Each produced DM state decays to a lighter DM state plus on- or off-shell $Z$. The lighter DM state in turn undergoes further decays, until the DM cascades down to a state so close to the gap that its decays are invisible in the detector due to its long lifetime and/or softness of its SM decay products. Overall, the event looks similar to the $X$+MET signature of neutralino $\tilde{\chi}^0_2$ production in SUSY models, but with higher multiplicity and somewhat softer spectrum of visible particles. 
Continuum DM production cross sections are suppressed by a factor of $\sin^4\alpha\sim 10^{-2}$ compared to SM electroweak processes, in addition to continuum kinematic suppression near the mass threshold similar to the suppression of direct detection rates. As a result, the Z-portal DM model is consistent with current LHC bounds. For $\mu_0<m_Z/2$, where $Z$-decays to on-shell DM pairs are allowed, strong bounds from LEP-1 and SLD become relevant. However, the bounds on WIC are again much weaker compared to the particle Z-portal DM, due to the continuum kinematic suppression of the decay rate. We conservatively estimate the bound by comparing the rate to the experimental limit on the invisible $Z$ decays $\Gamma_{\rm inv}$~\cite{Zyla:2020zbs}. The resulting bound is quite weak, and does not place a relevant constraint for $\rho_0=1$. For $\mu_0<m_h/2$ a Higgs decay to DM pairs would produce exotic semi-invisible final states. Demanding that no more than 10\% of Higgs bosons decay in this channel yields a bound indicated by the gray line in \cref{fig:Zportal}. We will explore collider phenomenology of the Z-portal WIC model in detail in future work.

\section{Conclusions}

In this letter we initiated the study of phenomenology of the Z-portal Weakly Interacting Continuum (WIC) model introduced in the companion paper~\cite{Csaki:2021gfm}. In this model the role of DM is played by a gapped continuum, rather than an ordinary particle, which acquires couplings to the SM $Z/W$ bosons from mixing with the Higgs. When the temperature is below the gap scale $\mu_0$, the WIC mass distribution is strongly peaked around $\mu_0$, and the thermal freeze-out of WICs is essentially identical to that of ordinary particle WIMPs. However other properties of the WIC are strikingly different. Due to the unique features of continuum kinematics, the direct detection cross section is strongly suppressed, re-opening the Z-portal for WICs. WIC decays $\Phi(\mu_1)\to \Phi(\mu_2)+$SM 
occur throughout the history of the universe, with each DM state decaying on average once per Hubble time. Such decays may yield new sources of $e^+e^-$ pairs during and after recombination, resulting in a {\it lower} bound on the WIC coupling to the SM. Collider signals of WICs include a cascade of SM particles and the invisible long-lived WIC states. Our analysis demonstrated that the Z-portal WIC provides a fully viable DM candidate consistent with all experimental constraints, as shown in the summary plot in \cref{fig:Zportal}. We look forward to further studies of the rich and novel phenomenology of these models.

\begin{acknowledgments}
	We thank Francesca Calore, Barry McCoy, Eun-Gook Moon, Carlos Wagner, Liantao Wang and Kathryn Zurek for useful discussions.
	C.C., S.H. G.K. and M.P. were supported in part by the NSF grant PHY-2014071. C.C. was also supported in part  by  the US-Israeli BSF grant 2016153.
	S.H.\  was also supported by the DOE grants DE-SC-0013642 and  DE-AC02-06CH11357, and by Hans Bethe Postdoctoral fellowship at Cornell.
	G.K. is supported by the Science and Technology Facilities Council with Grant No. ST/T000864/1.
	S.L.\ was supported by the Samsung Science and Technology Foundation under Project Number SSTF-BA1601-07 and by the National Research Foundation of Korea (NRF) grant funded by the Korea government (MEST) (No. NRF-2021R1A2C1005615).
	W.X. was supported in part by the DOE grant DE-SC0010296.
\end{acknowledgments}

\appendix

\section*{SUPPLEMENTARY MATERIAL}

\section{Out-of-Equilibrium Decays}
\label{app:Decay}

\begin{figure}[t!]
	\begin{center}
		\includegraphics[width=8cm]{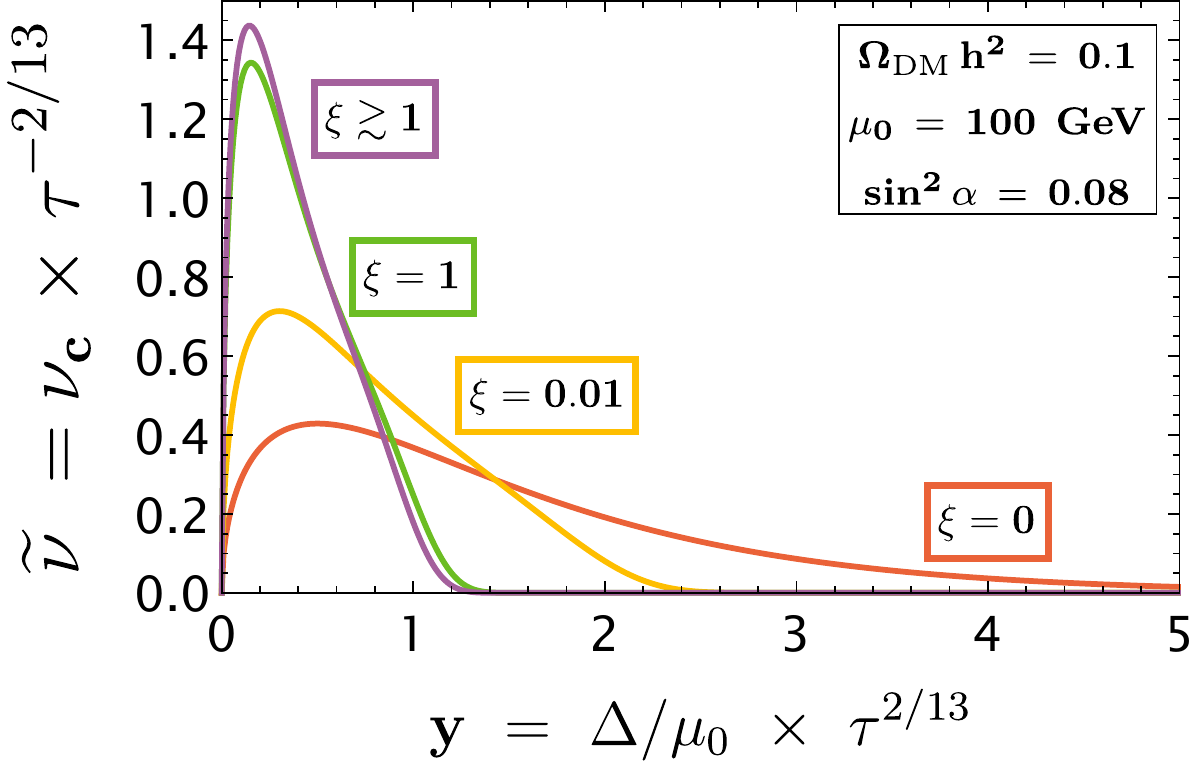} 
	\end{center}
	\caption{Evolution of mass distribution of DM states in rescaled variables $y$ and $\xi$. The Z-portal WIC model parameters are such that the relic density, CMB and direct detection constraints are satisfied.
	}
	\label{fig:decay_rescaled} 
\end{figure}

The mass distribution of continuum DM states after chemical decoupling follows a Boltzmann equation given by
\begin{align}
	\frac{\partial \nu(\mu^2)}{\partial t} + 3 H \nu(\mu^2) &\ = \ -\Gamma(\mu^2) \nu(\mu^2)\nonumber \\ 
	& + \int_{\mu^2}^{\infty} d\mu^{\prime\, 2} \nu(\mu^{\prime\, 2}) \frac{d\Gamma}{d\mu^2}(\mu^{\prime\, 2}\rightarrow \mu^2) 
	\label{eq:BoltzEq}
\end{align}
where the decay rate $\Gamma(\mu^2)$ is given in \cref{eq:decaytoee}, and
\beq
\frac{d\Gamma}{d\mu^2}(\mu^{\prime\, 2}\rightarrow \mu^2) \,=\, \frac{9009}{1024}\,\frac{\Gamma_0}{\mu_0^2}\,\left(\frac{\mu}{\mu_0}-1\right)^{1/2} \, (\mu^\prime-\mu)^5.
\eeq{dG}
The first term on the right-hand side of eq.~(\ref{eq:BoltzEq}) describes the depletion of states with mass $\mu$ due to their decays, while the second term represents their re-population due to decays of heavier modes. Note that 
\beq
\Gamma(\mu^{\prime 2})=\int_{\mu_0^2}^{\mu^{\prime 2}} d\mu^2 \,  \frac{d\Gamma}{d\mu^2}(\mu^{\prime\, 2}\rightarrow \mu^2).
\eeq{cons}
Using this relation, it is easy to check that the evolution described by \cref{eq:BoltzEq} conserves the total number of DM states, as expected due to $Z_2$ symmetry. 

The Boltzmann equation can be solved numerically to find the distribution of DM states at each redshift $z$. The numerical integration is smoother and more stable when $\mu$ and $t$ are rescaled as follows: $y = (\mu/\mu_0 - 1)\,  \tau^{2/13} $ and $\xi = \mathrm{log} (t/t_d)$, where $\tau = \Gamma_0 t$ and $t_d$ is the time at decoupling. (This rescaling is suggested by the estimate of the maximum DM mass using $\Gamma=H$, as described in the main text.) Additionally, we rescale the distribution $\nu$ as $\widetilde{\nu} = \nu_c \, \tau^{-2/13}$, where $\nu_c\,(\mu^2) = \nu a^3$ ($a=$ scale factor) is the comoving number density at each mass $\mu^2$. In the radiation dominated era, $a \propto t^{1/2}$ and $\nu_c = \nu\, (t/t_d)^{3/2}$. Ignoring overall normalization constants, an example of the evolution of this distribution is shown in \cref{fig:decay_rescaled}. The rescaled mass parameter remains of order one throughout the evolution. The initial evolution is fast, but the distribution stabilizes for $\xi \gtrsim 1$ as shown, and remains essentially the same for $\xi \gg 1$. Rescaling back to the original coordinates produces the distributions shown in \cref{fig:decay_log}.

\section{Electromagnetic Energy Injection After Recombination}

\begin{figure}[t!]
	\begin{center}
		\includegraphics[width=8cm]{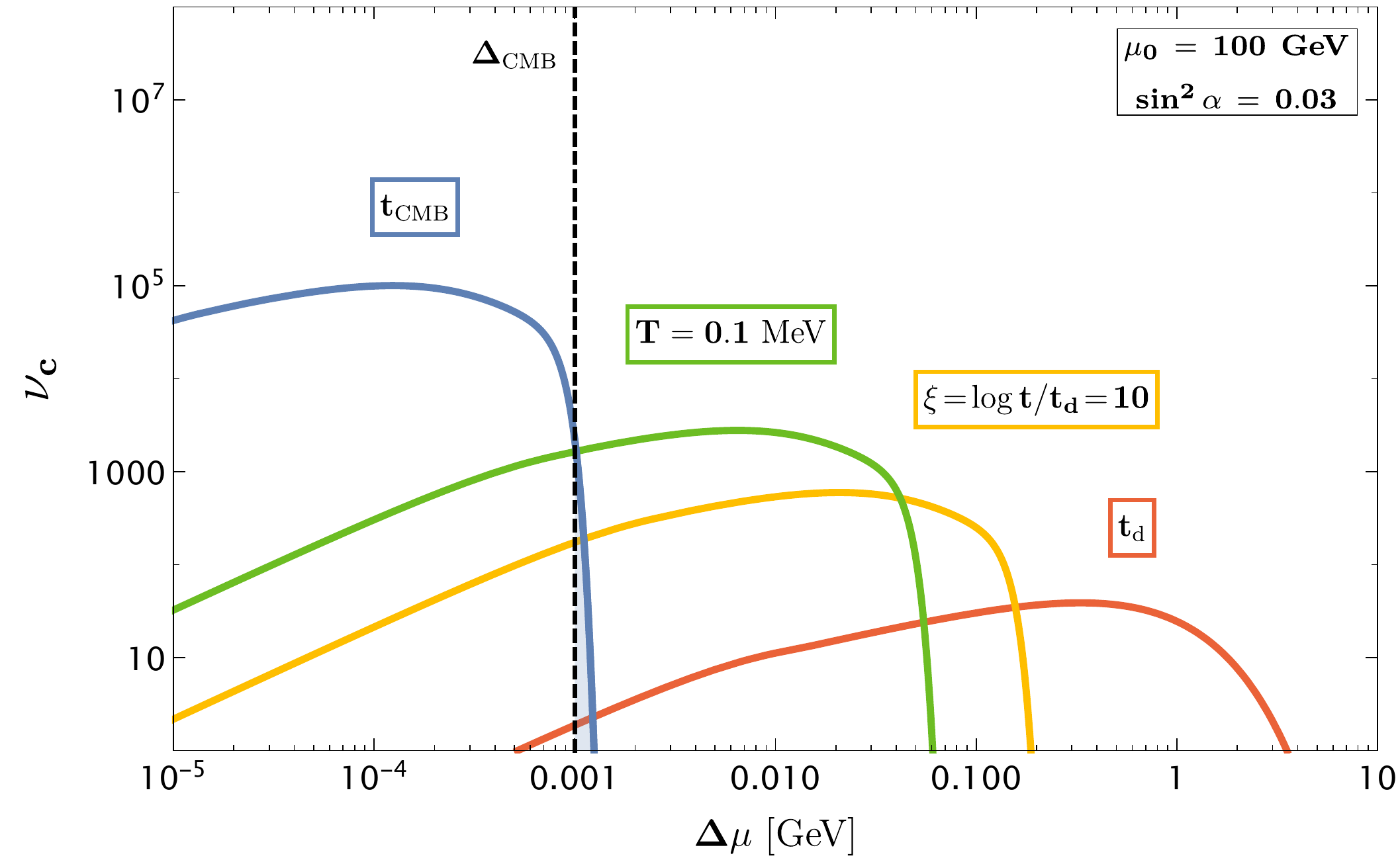} 
	\end{center}
	\caption{DM mass distribution as a function of time for a sample parameter point. The blue curve shows the distribution at the time of recombination; the dashed line corresponds to the kinematic threshold for the $e^+e^-$ final state in DM decay. Only about 0.2\% of DM particles lie on the tail above this threshold (the shaded area).   
	}
	\label{fig:dNdMsample} 
\end{figure}

One of the most important constraints on the model arises from re-ionization of Hydrogen by DM decays after recombination. In \cref{fig:Zportal}, this bound is estimated using a simple criterion $\Gamma=H(t_{\rm CMB})$. To confirm the validity of this estimate, we performed a more detailed analysis of this process. Hydrogen can only be ionized by the decays of DM particles with masses above the kinematic threshold to produce $e^+e^-$ pairs. The fraction of ionized Hydrogen atoms is given by
\beqa
r_{\rm ion} &=& \left( \frac{N_{DM}}{N_B}\right) \, \int_{\mu_0+2m_e}^{\infty}d\mu\,\left(\frac{dN_{DM}}{d\mu}\right)\,\int_{\mu_0}^{\mu-2m_e} d\mu^\prime \,\CR & & \hskip-1cm \times \frac{\rho(\mu^\prime)}{\Gamma_0(\mu)}\,\frac{d\Gamma(\Phi(\mu)\to \Phi(\mu^\prime) e^+e^-)}{d\mu^\prime} \,\left(\frac{K_e(\mu, \mu^\prime)}{E_{\rm ion}}\right).
\eeqa{ionize} 	
Here $dN_{DM}/d\mu$ is the mass distribution of DM at the time of recombination; $\frac{N_{DM}}{N_B} = \frac{\Omega_{DM}}{\Omega_B}\times \frac{m_p}{m_{DM}} \approx 5 \frac{m_p}{\mu_0}$; $E_{\rm ion}=13.6$~eV is the H ionization energy; and $K_e$ is the average (total) electron kinetic energy produced in the decay. For simplicity, we conservatively assumed a 100\% ionization efficiency. 
Since the decaying DM is non-relativistic, conservation of energy requires $K_e \leq \mu - \mu^\prime - 2m_e$. Conservatively, we set $K_e = \mu-\mu_0-2m_e$ for all $\mu^\prime$. Furthermore, $\Gamma_0(\mu)$ is the total decay width of the $\Phi$, dominated by decays into $\nu\bar{\nu}$. For DM mass far above threshold for $e^+e^-$ decays, we have 
\beq
\int_{\mu_0}^{\mu-2m_e} d\mu^\prime \,\frac{\rho(\mu^\prime)}{\Gamma_0(\mu)}\,\frac{d\Gamma(\Phi(\mu)\to \Phi(\mu^\prime) e^+e^-)}{d\mu^\prime} \approx \frac{g_{Ze}^2}{3g_{Z\nu}^2} \approx \frac{1}{6}. 
\eeq{Br}
Near threshold there is additional suppression that further reduces this branching ratio. First, there is phase-space suppression in the decay rate for $\Phi(\mu)\to \Phi(\mu^\prime) e^+e^-$. 
In addition, there is suppression from the narrow domain of $\mu^\prime$ integration, whose range is $\mu-\mu_0 \sim 1$~MeV for neutrino decays but only $\sim \mu-\mu_0-2m_e$ for electron decays. 
Finally, there is extra suppression due to smallness of spectral density $\rho$ near the gap, which is stronger in the electron rate compared to neutrinos since $\mu^\prime$ is on average closer to the gap in the former case. Overall, the branching ratio is suppressed by at least a factor of $(K_e/2m_e)^{1.5}$.    

The mass distribution $\frac{dN_{DM}}{d\mu}$ is obtained by numerical integration of the Boltzmann equation. For example, consider a sample point with parameters $(\mu_0=100$~GeV$, \sin^2\alpha=0.03)$, which lies on the boundary of the allowed region in Fig.~\ref{fig:Zportal}. The mass distribution at this point at the time of recombination is shown in Fig.~\ref{fig:dNdMsample}. Only about 0.2\% of DM particles have masses above threshold for $e^+e^-$ decays. Using this distribution in Eq.~\leqn{ionize}, we find a conservative bound $r_{\rm ion}\lsim 10^{-3}$. At the same time, CMB observations require that at most about 1\% of Hydrogen can be re-ionized after CMB decoupling~\cite{Slatyer:2016qyl,Poulin:2016anj}. Thus we conclude that this sample point is in fact consistent with the CMB constraint. Repeating the analysis for a sample of points on the boundary of the allowed region in fig.~\ref{fig:Zportal} yields the same conclusion, so the CMB bound shown in the figure is in fact overly conservative. However, the EM energy injected by DM decays changes exponentially fast with the model parameters around the boundary. This is because only DM particles that survived longer than their nominal lifetime can decay in the $e^+e^-$ channel, and the fraction of such particles is exponentially suppressed with shorter nominal lifetime. As a result, the more careful analysis yields a very slight weakening of the CMB bound compared to the simple estimate $\Gamma=H(t_{\rm CMB})$, but does not significantly change our discussion of WIC DM parameter space.            

\section{Effects of Varying $\rho_0$}

\begin{figure}
	\begin{center}
		\includegraphics[width=9.5cm]{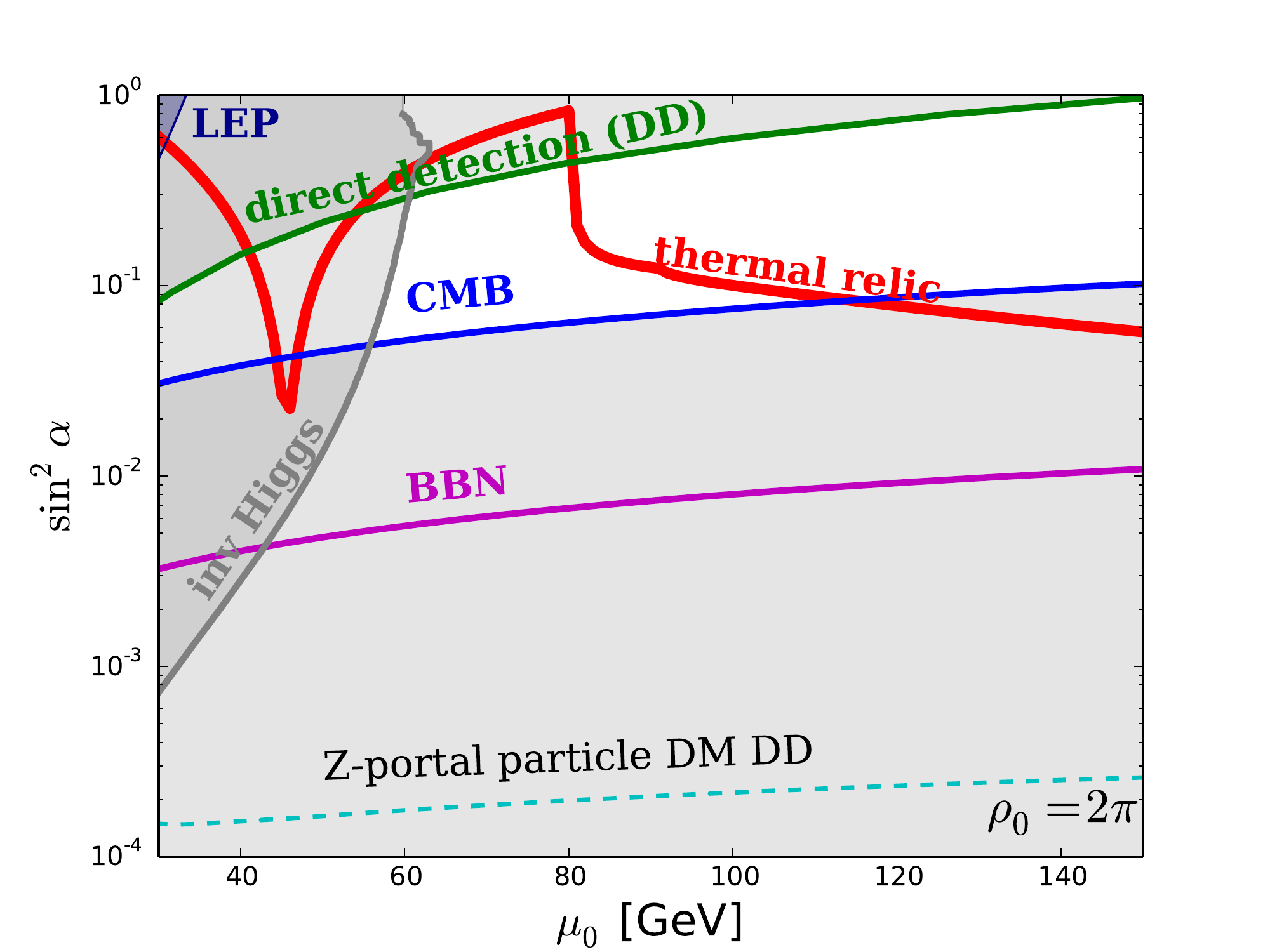} 
	\end{center}
	\caption{Parameter space of the Z-portal WIC, with $\rho_0=2\pi$. Color scheme is the same as in \cref{fig:Zportal}, with the addition of the direct detection constraint (red line) and LEP bound (dark-gray line) which become relevant due to enhanced effects of WIC.  
	}
	\label{fig:Zportal2pi} 
\end{figure}

The WIC model contains an additional parameter, $\rho_0$, defined in \cref{eq:rho0}. We chose $\rho_0=1$ as a benchmark value in the discussion above, including the summary plot \cref{fig:Zportal}. Here we illustrate the effect of changing $\rho_0$ by presenting the version of the summary plot with $\rho_0=2\pi$; see \cref{fig:Zportal2pi}. Increasing $\rho_0$ increases most of the phenomenologically releveant cross sections and decay rates, shifting the allowed band in the parameter space to lower couplings. However, the region with correct relic density is not significantly affected. As a result, the viable parameter space shifts to larger values of the gap scale, $80-110$~GeV in this case vs. $60-80$ GeV for $\rho_0=1$.  

\bibliographystyle{utphys}
\bibliography{ref}

\providecommand{\href}[2]{#2}\begingroup\raggedright\begin{thebibliography}{10}

\bibitem{Lee:1977ua}
B.~W. Lee and S.~Weinberg, ``{Cosmological Lower Bound on Heavy Neutrino
  Masses},'' \href{http://dx.doi.org/10.1103/PhysRevLett.39.165}{{\em Phys.
  Rev. Lett.} {\bfseries 39} (1977) 165--168}.

\bibitem{Jungman:1995df}
G.~Jungman, M.~Kamionkowski, and K.~Griest, ``{Supersymmetric dark matter},''
  \href{http://dx.doi.org/10.1016/0370-1573(95)00058-5}{{\em Phys. Rept.}
  {\bfseries 267} (1996) 195--373},
  \href{http://arxiv.org/abs/hep-ph/9506380}{{\ttfamily arXiv:hep-ph/9506380}}.

\bibitem{Bertone:2004pz}
G.~Bertone, D.~Hooper, and J.~Silk, ``{Particle dark matter: Evidence,
  candidates and constraints},''
  \href{http://dx.doi.org/10.1016/j.physrep.2004.08.031}{{\em Phys. Rept.}
  {\bfseries 405} (2005) 279--390},
  \href{http://arxiv.org/abs/hep-ph/0404175}{{\ttfamily arXiv:hep-ph/0404175}}.

\bibitem{Escudero:2016gzx}
M.~Escudero, A.~Berlin, D.~Hooper, and M.-X. Lin, ``{Toward (Finally!) Ruling
  Out Z and Higgs Mediated Dark Matter Models},''
  \href{http://dx.doi.org/10.1088/1475-7516/2016/12/029}{{\em JCAP} {\bfseries
  12} (2016) 029}, \href{http://arxiv.org/abs/1609.09079}{{\ttfamily
  arXiv:1609.09079 [hep-ph]}}.

\bibitem{Aprile:2018dbl}
{\bfseries XENON} Collaboration, E.~Aprile {\em et~al.}, ``{Dark Matter Search
  Results from a One Ton-Year Exposure of XENON1T},''
  \href{http://dx.doi.org/10.1103/PhysRevLett.121.111302}{{\em Phys. Rev.
  Lett.} {\bfseries 121} no.~11, (2018) 111302},
  \href{http://arxiv.org/abs/1805.12562}{{\ttfamily arXiv:1805.12562
  [astro-ph.CO]}}.

\bibitem{Akerib:2016vxi}
{\bfseries LUX} Collaboration, D.~S. Akerib {\em et~al.}, ``{Results from a
  search for dark matter in the complete LUX exposure},''
  \href{http://dx.doi.org/10.1103/PhysRevLett.118.021303}{{\em Phys. Rev.
  Lett.} {\bfseries 118} no.~2, (2017) 021303},
  \href{http://arxiv.org/abs/1608.07648}{{\ttfamily arXiv:1608.07648
  [astro-ph.CO]}}.

\bibitem{Cui:2017nnn}
{\bfseries PandaX-II} Collaboration, X.~Cui {\em et~al.}, ``{Dark Matter
  Results From 54-Ton-Day Exposure of PandaX-II Experiment},''
  \href{http://dx.doi.org/10.1103/PhysRevLett.119.181302}{{\em Phys. Rev.
  Lett.} {\bfseries 119} no.~18, (2017) 181302},
  \href{http://arxiv.org/abs/1708.06917}{{\ttfamily arXiv:1708.06917
  [astro-ph.CO]}}.

\bibitem{Ade:2015xua}
{\bfseries Planck} Collaboration, P.~A.~R. Ade {\em et~al.}, ``{Planck 2015
  results. XIII. Cosmological parameters},''
  \href{http://dx.doi.org/10.1051/0004-6361/201525830}{{\em Astron. Astrophys.}
  {\bfseries 594} (2016) A13},
  \href{http://arxiv.org/abs/1502.01589}{{\ttfamily arXiv:1502.01589
  [astro-ph.CO]}}.

\bibitem{Slatyer:2016qyl}
T.~R. Slatyer and C.-L. Wu, ``{General Constraints on Dark Matter Decay from
  the Cosmic Microwave Background},''
  \href{http://dx.doi.org/10.1103/PhysRevD.95.023010}{{\em Phys. Rev. D}
  {\bfseries 95} no.~2, (2017) 023010},
  \href{http://arxiv.org/abs/1610.06933}{{\ttfamily arXiv:1610.06933
  [astro-ph.CO]}}.

\bibitem{Slatyer:2015jla}
T.~R. Slatyer, ``{Indirect dark matter signatures in the cosmic dark ages. I.
  Generalizing the bound on s-wave dark matter annihilation from Planck
  results},'' \href{http://dx.doi.org/10.1103/PhysRevD.93.023527}{{\em Phys.
  Rev. D} {\bfseries 93} no.~2, (2016) 023527},
  \href{http://arxiv.org/abs/1506.03811}{{\ttfamily arXiv:1506.03811
  [hep-ph]}}.

\bibitem{Reno:1987qw}
M.~H. Reno and D.~Seckel, ``{Primordial Nucleosynthesis: The Effects of
  Injecting Hadrons},'' \href{http://dx.doi.org/10.1103/PhysRevD.37.3441}{{\em
  Phys. Rev. D} {\bfseries 37} (1988) 3441}.

\bibitem{Kawasaki:1994sc}
M.~Kawasaki and T.~Moroi, ``{Electromagnetic cascade in the early universe and
  its application to the big bang nucleosynthesis},''
  \href{http://dx.doi.org/10.1086/176324}{{\em Astrophys. J.} {\bfseries 452}
  (1995) 506}, \href{http://arxiv.org/abs/astro-ph/9412055}{{\ttfamily
  arXiv:astro-ph/9412055}}.

\bibitem{Cyburt:2002uv}
R.~H. Cyburt, J.~R. Ellis, B.~D. Fields, and K.~A. Olive, ``{Updated
  nucleosynthesis constraints on unstable relic particles},''
  \href{http://dx.doi.org/10.1103/PhysRevD.67.103521}{{\em Phys. Rev. D}
  {\bfseries 67} (2003) 103521},
  \href{http://arxiv.org/abs/astro-ph/0211258}{{\ttfamily
  arXiv:astro-ph/0211258}}.

\bibitem{Kawasaki:2004qu}
M.~Kawasaki, K.~Kohri, and T.~Moroi, ``{Big-Bang nucleosynthesis and hadronic
  decay of long-lived massive particles},''
  \href{http://dx.doi.org/10.1103/PhysRevD.71.083502}{{\em Phys. Rev. D}
  {\bfseries 71} (2005) 083502},
  \href{http://arxiv.org/abs/astro-ph/0408426}{{\ttfamily
  arXiv:astro-ph/0408426}}.

\bibitem{Hisano:2009rc}
J.~Hisano, M.~Kawasaki, K.~Kohri, T.~Moroi, and K.~Nakayama, ``{Cosmic Rays
  from Dark Matter Annihilation and Big-Bang Nucleosynthesis},''
  \href{http://dx.doi.org/10.1103/PhysRevD.79.083522}{{\em Phys. Rev. D}
  {\bfseries 79} (2009) 083522},
  \href{http://arxiv.org/abs/0901.3582}{{\ttfamily arXiv:0901.3582 [hep-ph]}}.

\bibitem{Henning:2012rm}
B.~Henning and H.~Murayama, ``{Constraints on Light Dark Matter from Big Bang
  Nucleosynthesis},'' \href{http://arxiv.org/abs/1205.6479}{{\ttfamily
  arXiv:1205.6479 [hep-ph]}}.

\bibitem{Csaki:2021gfm}
C.~Cs\'aki, S.~Hong, G.~Kurup, S.~J. Lee, M.~Perelstein, and W.~Xue,
  ``{Continuum Dark Matter},''
\href{http://arxiv.org/abs/2105.07035}{{\ttfamily arXiv:2105.07035 [hep-ph]}}.

\bibitem{Dienes:2011ja}
K.~R. Dienes and B.~Thomas, ``{Dynamical Dark Matter: I. Theoretical
  Overview},'' \href{http://dx.doi.org/10.1103/PhysRevD.85.083523}{{\em Phys.
  Rev. D} {\bfseries 85} (2012) 083523},
  \href{http://arxiv.org/abs/1106.4546}{{\ttfamily arXiv:1106.4546 [hep-ph]}}.

\bibitem{Dienes:2011sa}
K.~R. Dienes and B.~Thomas, ``{Dynamical Dark Matter: II. An Explicit Model},''
  \href{http://dx.doi.org/10.1103/PhysRevD.85.083524}{{\em Phys. Rev. D}
  {\bfseries 85} (2012) 083524},
  \href{http://arxiv.org/abs/1107.0721}{{\ttfamily arXiv:1107.0721 [hep-ph]}}.

\bibitem{Dienes:2012yz}
K.~R. Dienes, S.~Su, and B.~Thomas, ``{Distinguishing Dynamical Dark Matter at
  the LHC},'' \href{http://dx.doi.org/10.1103/PhysRevD.86.054008}{{\em Phys.
  Rev. D} {\bfseries 86} (2012) 054008},
  \href{http://arxiv.org/abs/1204.4183}{{\ttfamily arXiv:1204.4183 [hep-ph]}}.

\bibitem{Dienes:2012cf}
K.~R. Dienes, J.~Kumar, and B.~Thomas, ``{Direct Detection of Dynamical Dark
  Matter},'' \href{http://dx.doi.org/10.1103/PhysRevD.86.055016}{{\em Phys.
  Rev. D} {\bfseries 86} (2012) 055016},
  \href{http://arxiv.org/abs/1208.0336}{{\ttfamily arXiv:1208.0336 [hep-ph]}}.

\bibitem{Dienes:2013xya}
K.~R. Dienes, J.~Kumar, B.~Thomas, and D.~Yaylali, ``{Overcoming Velocity
  Suppression in Dark-Matter Direct-Detection Experiments},''
  \href{http://dx.doi.org/10.1103/PhysRevD.90.015012}{{\em Phys. Rev. D}
  {\bfseries 90} no.~1, (2014) 015012},
  \href{http://arxiv.org/abs/1312.7772}{{\ttfamily arXiv:1312.7772 [hep-ph]}}.

\bibitem{Dienes:2014via}
K.~R. Dienes, J.~Kumar, B.~Thomas, and D.~Yaylali, ``{Dark-Matter Decay as a
  Complementary Probe of Multicomponent Dark Sectors},''
  \href{http://dx.doi.org/10.1103/PhysRevLett.114.051301}{{\em Phys. Rev.
  Lett.} {\bfseries 114} no.~5, (2015) 051301},
  \href{http://arxiv.org/abs/1406.4868}{{\ttfamily arXiv:1406.4868 [hep-ph]}}.

\bibitem{Dienes:2014bka}
K.~R. Dienes, S.~Su, and B.~Thomas, ``{Strategies for probing nonminimal dark
  sectors at colliders: The interplay between cuts and kinematic
  distributions},'' \href{http://dx.doi.org/10.1103/PhysRevD.91.054002}{{\em
  Phys. Rev. D} {\bfseries 91} no.~5, (2015) 054002},
  \href{http://arxiv.org/abs/1407.2606}{{\ttfamily arXiv:1407.2606 [hep-ph]}}.

\bibitem{Boddy:2016fds}
K.~K. Boddy, K.~R. Dienes, D.~Kim, J.~Kumar, J.-C. Park, and B.~Thomas,
  ``{Lines and Boxes: Unmasking Dynamical Dark Matter through Correlations in
  the MeV Gamma-Ray Spectrum},''
  \href{http://dx.doi.org/10.1103/PhysRevD.94.095027}{{\em Phys. Rev. D}
  {\bfseries 94} no.~9, (2016) 095027},
  \href{http://arxiv.org/abs/1606.07440}{{\ttfamily arXiv:1606.07440
  [hep-ph]}}.

\bibitem{Curtin:2018ees}
D.~Curtin, K.~R. Dienes, and B.~Thomas, ``{Dynamical Dark Matter, MATHUSLA, and
  the Lifetime Frontier},''
  \href{http://dx.doi.org/10.1103/PhysRevD.98.115005}{{\em Phys. Rev. D}
  {\bfseries 98} no.~11, (2018) 115005},
  \href{http://arxiv.org/abs/1809.11021}{{\ttfamily arXiv:1809.11021
  [hep-ph]}}.

\bibitem{Dienes:2019krh}
K.~R. Dienes, D.~Kim, H.~Song, S.~Su, B.~Thomas, and D.~Yaylali, ``{Nonminimal
  dark sectors: Mediator-induced decay chains and multijet collider
  signatures},'' \href{http://dx.doi.org/10.1103/PhysRevD.101.075024}{{\em
  Phys. Rev. D} {\bfseries 101} no.~7, (2020) 075024},
  \href{http://arxiv.org/abs/1910.01129}{{\ttfamily arXiv:1910.01129
  [hep-ph]}}.

\bibitem{vonHarling:2012sz}
B.~von Harling and K.~L. McDonald, ``{Secluded Dark Matter Coupled to a Hidden
  CFT},'' \href{http://dx.doi.org/10.1007/JHEP08(2012)048}{{\em JHEP}
  {\bfseries 08} (2012) 048}, \href{http://arxiv.org/abs/1203.6646}{{\ttfamily
  arXiv:1203.6646 [hep-ph]}}.

\bibitem{Katz:2015zba}
A.~Katz, M.~Reece, and A.~Sajjad, ``{Continuum-mediated dark
  matter\textendash{}baryon scattering},''
  \href{http://dx.doi.org/10.1016/j.dark.2016.01.002}{{\em Phys. Dark Univ.}
  {\bfseries 12} (2016) 24--36},
  \href{http://arxiv.org/abs/1509.03628}{{\ttfamily arXiv:1509.03628
  [hep-ph]}}.

\bibitem{Chaffey:2021tmj}
I.~Chaffey, S.~Fichet, and P.~Tanedo, ``{Continuum-Mediated Self-Interacting
  Dark Matter},'' \href{http://dx.doi.org/10.1007/JHEP06(2021)008}{{\em JHEP}
  {\bfseries 06} (2021) 008}, \href{http://arxiv.org/abs/2102.05674}{{\ttfamily
  arXiv:2102.05674 [hep-ph]}}.

\bibitem{Georgi:2007ek}
H.~Georgi, ``{Unparticle physics},''
  \href{http://dx.doi.org/10.1103/PhysRevLett.98.221601}{{\em Phys. Rev. Lett.}
  {\bfseries 98} (2007) 221601},
  \href{http://arxiv.org/abs/hep-ph/0703260}{{\ttfamily arXiv:hep-ph/0703260}}.

\bibitem{Stancato:2008mp}
D.~Stancato and J.~Terning, ``{The Unhiggs},''
  \href{http://dx.doi.org/10.1088/1126-6708/2009/11/101}{{\em JHEP} {\bfseries
  11} (2009) 101},
\href{http://arxiv.org/abs/0807.3961}{{\ttfamily arXiv:0807.3961 [hep-ph]}}.

\bibitem{Falkowski:2008yr}
A.~Falkowski and M.~Perez-Victoria, ``{Holographic Unhiggs},''
  \href{http://dx.doi.org/10.1103/PhysRevD.79.035005}{{\em Phys. Rev.}
  {\bfseries D79} (2009) 035005},
\href{http://arxiv.org/abs/0810.4940}{{\ttfamily arXiv:0810.4940 [hep-ph]}}.

\bibitem{Bellazzini:2015cgj}
B.~Bellazzini, C.~Cs\'aki, J.~Hubisz, S.~J. Lee, J.~Serra, and J.~Terning,
  ``{Quantum Critical Higgs},''
  \href{http://dx.doi.org/10.1103/PhysRevX.6.041050}{{\em Phys. Rev. X}
  {\bfseries 6} no.~4, (2016) 041050},
  \href{http://arxiv.org/abs/1511.08218}{{\ttfamily arXiv:1511.08218
  [hep-ph]}}.

\bibitem{Csaki:2018kxb}
C.~Cs\'aki, G.~Lee, S.~J. Lee, S.~Lombardo, and O.~Telem, ``{Continuum
  Naturalness},'' \href{http://dx.doi.org/10.1007/JHEP03(2019)142}{{\em JHEP}
  {\bfseries 03} (2019) 142}, \href{http://arxiv.org/abs/1811.06019}{{\ttfamily
  arXiv:1811.06019 [hep-ph]}}.

\bibitem{Gubser:2000nd}
S.~S. Gubser, ``{Curvature singularities: The Good, the bad, and the naked},''
  \href{http://dx.doi.org/10.4310/ATMP.2000.v4.n3.a6}{{\em Adv. Theor. Math.
  Phys.} {\bfseries 4} (2000) 679--745},
  \href{http://arxiv.org/abs/hep-th/0002160}{{\ttfamily arXiv:hep-th/0002160}}.

\bibitem{sachdev2007quantum}
S.~Sachdev, ``Quantum phase transitions,'' {\em Handbook of Magnetism and
  Advanced Magnetic Materials} (2007) .

\bibitem{Fradkin:1991nr}
E.~H. Fradkin, {\em {Field Theories of Condensed Matter Physics}}, vol.~82.
\newblock Cambridge Univ. Press, Cambridge, UK, 2, 2013.

\bibitem{McCoy:1978ta}
B.~M. McCoy and T.~T. Wu, ``{Two-dimensional Ising Field Theory in a Magnetic
  Field: Breakup of the Cut in the Two Point Function},''
  \href{http://dx.doi.org/10.1103/PhysRevD.18.1259}{{\em Phys. Rev. D}
  {\bfseries 18} (1978) 1259}.

\bibitem{Randall:1999ee}
L.~Randall and R.~Sundrum, ``{A Large mass hierarchy from a small extra
  dimension},'' \href{http://dx.doi.org/10.1103/PhysRevLett.83.3370}{{\em Phys.
  Rev. Lett.} {\bfseries 83} (1999) 3370--3373},
  \href{http://arxiv.org/abs/hep-ph/9905221}{{\ttfamily arXiv:hep-ph/9905221}}.

\bibitem{Cabrer:2009we}
J.~A. Cabrer, G.~von Gersdorff, and M.~Quiros, ``{Soft-Wall Stabilization},''
  \href{http://dx.doi.org/10.1088/1367-2630/12/7/075012}{{\em New J. Phys.}
  {\bfseries 12} (2010) 075012},
  \href{http://arxiv.org/abs/0907.5361}{{\ttfamily arXiv:0907.5361 [hep-ph]}}.

\bibitem{Aprile:2020tmw}
{\bfseries XENON} Collaboration, E.~Aprile {\em et~al.}, ``{Excess electronic
  recoil events in XENON1T},''
  \href{http://dx.doi.org/10.1103/PhysRevD.102.072004}{{\em Phys. Rev. D}
  {\bfseries 102} no.~7, (2020) 072004},
  \href{http://arxiv.org/abs/2006.09721}{{\ttfamily arXiv:2006.09721
  [hep-ex]}}.

\bibitem{Bays:2011si}
{\bfseries Super-Kamiokande} Collaboration, K.~Bays {\em et~al.}, ``{Supernova
  Relic Neutrino Search at Super-Kamiokande},''
  \href{http://dx.doi.org/10.1103/PhysRevD.85.052007}{{\em Phys. Rev. D}
  {\bfseries 85} (2012) 052007},
  \href{http://arxiv.org/abs/1111.5031}{{\ttfamily arXiv:1111.5031 [hep-ex]}}.

\bibitem{Agostini:2020lci}
{\bfseries BOREXINO} Collaboration, M.~Agostini {\em et~al.}, ``{Sensitivity to
  neutrinos from the solar CNO cycle in Borexino},''
  \href{http://dx.doi.org/10.1140/epjc/s10052-020-08534-2}{{\em Eur. Phys. J.
  C} {\bfseries 80} no.~11, (2020) 1091},
  \href{http://arxiv.org/abs/2005.12829}{{\ttfamily arXiv:2005.12829
  [hep-ex]}}.

\bibitem{Agostini:2020mfq}
{\bfseries BOREXINO} Collaboration, M.~Agostini {\em et~al.}, ``{Experimental
  evidence of neutrinos produced in the CNO fusion cycle in the Sun},''
  \href{http://dx.doi.org/10.1038/s41586-020-2934-0}{{\em Nature} {\bfseries
  587} (2020) 577--582}, \href{http://arxiv.org/abs/2006.15115}{{\ttfamily
  arXiv:2006.15115 [hep-ex]}}.

\bibitem{Lewin:1995rx}
J.~D. Lewin and P.~F. Smith, ``{Review of mathematics, numerical factors, and
  corrections for dark matter experiments based on elastic nuclear recoil},''
  \href{http://dx.doi.org/10.1016/S0927-6505(96)00047-3}{{\em Astropart. Phys.}
  {\bfseries 6} (1996) 87--112}.

\bibitem{Fermi-LAT:2016uux}
{\bfseries Fermi-LAT, DES} Collaboration, A.~Albert {\em et~al.}, ``{Searching
  for Dark Matter Annihilation in Recently Discovered Milky Way Satellites with
  Fermi-LAT},'' \href{http://dx.doi.org/10.3847/1538-4357/834/2/110}{{\em
  Astrophys. J.} {\bfseries 834} no.~2, (2017) 110},
  \href{http://arxiv.org/abs/1611.03184}{{\ttfamily arXiv:1611.03184
  [astro-ph.HE]}}.

\bibitem{Zyla:2020zbs}
{\bfseries Particle Data Group} Collaboration, P.~Zyla {\em et~al.}, ``{Review
  of Particle Physics},'' \href{http://dx.doi.org/10.1093/ptep/ptaa104}{{\em
  PTEP} {\bfseries 2020} no.~8, (2020) 083C01}.

\bibitem{Poulin:2016anj}
V.~Poulin, J.~Lesgourgues, and P.~D. Serpico, ``{Cosmological constraints on
  exotic injection of electromagnetic energy},''
  \href{http://dx.doi.org/10.1088/1475-7516/2017/03/043}{{\em JCAP} {\bfseries
  03} (2017) 043}, \href{http://arxiv.org/abs/1610.10051}{{\ttfamily
  arXiv:1610.10051 [astro-ph.CO]}}.

\end{thebibliography}\endgroup

\end{document}